\documentclass[twocolumn,twoside,slac_two]{revtex4}
\usepackage{graphicx}
\usepackage{fancyhdr}
\begin{document}

\title{Underground Searches for Cold Relics of the Early Universe}

\author{Laura Baudis}
\affiliation{University of Florida, Gainesville, FL 32611, USA}

\begin{abstract}

We have strong evidence on all cosmic scales, from galaxies to the largest structures ever observed, 
that there is more matter in the universe than we can see. Galaxies and clusters would fly apart unless they
would be held together by material which we call dark, because it does not shine in photons.
Although the amount of dark matter and its distribution are fairly well established, we are clueless 
regarding its  composition. Leading candidates are Weakly Interacting Massive Particles (WIMPs), which are
'cold' thermal relics  of the Big Bang, ie moving non-relativistically at the time of structure formation. 
These particles can be detected via their interaction with nuclei in deep-underground, low-background detectors.
Experiments dedicated to observe WIMP interactions for the first time reach sensitivities allowing to probe  
the parameter space predicted by supersymmetric theories of particle physics. 
Current results of high sensitivity direct detection experiments are discussed and the most promising projects 
of the future are presented. If a stable new particle exists at the weak scale, it seems likely 
to expect a discovery within this decade.

\end{abstract}

\maketitle
\thispagestyle{fancy}

\section{Introduction}

More than seventy years after Zwicky's first accounts of dark matter in galaxy clusters \cite{zwicky33}, 
and thirty five years after Rubin's measurements of rotational velocities of spirals \cite{rubin70},
the case for non-baryonic dark matter remains convincing. Recent precision observations of the 
cosmic microwave background \cite{bennet03} and of large scale structures \cite{sdss03} confirm the picture in 
which more than 90\% of the matter in the universe is revealed only by its gravitational interaction.
The nature of this matter is not known. A class of generic candidates are weakly interacting massive 
particles (WIMPs) which could have been thermally produced in the very early universe. 
It is well known that if the mass and cross section of these particles is determined by the weak scale,
the freeze-out relic density is around the observed value, $\Omega \sim$ 0.1.

The prototype WIMP candidate is the neutralino, or the lightest supersymmetric particle, which 
is stable in supersymmetric models where R-parity is conserved.  Another recently discussed candidate 
is the lightest Kaluza-Klein excitation (LKP) in theories with universal extra dimensions. If a new 
discrete symmetry, called KK-parity is conserved, and if the KK particle masses are related 
to the weak scale, the LKP is stable and makes an excellent dark matter candidate.
A vast experimental effort to detect WIMPs is underway.  
For excellent recent reviews we refer to \cite{rick04}, \cite{gabriel05} and \cite{bertone04}.
The good news is that cryogenic experiments are 
now for the first time probing the parameter space predicted by SUSY theories for neutralinos.
On the more pessimistic side, predicted WIMP quark cross sections span several orders of magnitude, and ton 
or multi-ton scale detectors may be required for a detection. However, such experiments are now in the 
stage of development, with prototypes being tested and installed in underground labs. The challenge is immense, 
but the rewards would be outstanding: revealing the major constituents of matter in the universe and their doubtless 
profound implications for fundamental physics at the weak scale.

\section{\label{directdet}Direct Detection of WIMPs}

WIMPs can be detected directly, via their scattering off nuclei in terrestrial targets \cite{goodman85}, 
or indirectly, via their annihilation products in the Sun, Earth, galactic halo and galactic center with neutrino 
telescopes and space-based detectors. Here we will briefly discuss direct detection only. 

The differential rate for WIMP elastic scattering off nuclei is given by 
\cite{smith_lewin96}

\begin{equation}
\frac {dR}{dE_R}=N_{T}
\frac{\rho_{0}}{m_{W}}
                    \int_{v_{\rm min}}^{v_{\rm max}} \,d \vec{v}\,f(\vec v)\,v
                     \,\frac{d\sigma}{d E_R}\, , 
\label{eq1}
\end{equation}

where $N_T$ represents the number of the target nuclei,
$m_W$ is the WIMP mass and $\rho_0$
the local WIMP density in the galactic halo,
$\vec v$ and $f(\vec v)$ are the WIMP
velocity and velocity distribution function  in the Earth frame
 and ${d\sigma}/{d E_R}$ is the WIMP-nucleus differential cross section.

The nuclear recoil energy is given by
$E_R={{m_{\rm r}^2}}v^2(1-\cos \theta)/{m_N}$,
where $\theta$ is the  scattering
angle in the WIMP-nucleus center-of-mass frame,
$m_N$ is the nuclear mass and $m_{\rm r}$ is the WIMP-nucleus
reduced mass. The velocity $v_{\rm min}$ is defined as 
$v_{\rm min} = (m_N E_{th}/2m_{\rm r}^2)^{\frac{1}{2}}$, where $E_{th}$
is the energy threshold of the detector, and 
$v_{\rm max}$ is the escape WIMP velocity in the Earth frame.  

The simplest galactic model assumes a Maxwell-Boltzmann distribution for the
WIMP velocity in the galactic rest frame, with a velocity dispersion 
of v$_{rms} \approx$ 270 km s$^{-1}$ and an escape 
velocity of v$_{esc} \approx$ 650 km s$^{-1}$. 

The differential WIMP-nucleus cross section can have two 
separate components: an effective scalar coupling between the WIMP and the 
nucleus (proportional to A$^2$, where A is the target atomic mass) and 
an effective coupling between the spin of the WIMP and the total spin of the nucleus.
In general the coherent part dominates the interaction (depending however 
on the content of the neutralino) 
and the cross section  can be factorized as 
$\frac{d\sigma}{d E_R} \propto {\sigma_0} F^2(E_R)$, where 
 $\sigma_0$ is the point-like scalar WIMP-nucleus
cross section and $F(E_R)$ denotes the nuclear form factor, 
expressed as a function of the recoil energy.

The left side of equation \ref{eq1} is the measured spectrum in a detector,
while the right side represents the theoretical prediction. 
It includes  WIMP properties which are completely unknown,
such as the WIMP mass $m_W$ and elastic cross section $\sigma_0$, 
quantities accessible from astrophysics, such as the density of WIMPs in the 
halo, $\rho_0$, the WIMP velocity distribution and the escape
velocity (which are however prone to large uncertainties) and
detector specific parameters: mass of target nucleus, energy
threshold and nuclear form factor.

\begin{figure}[thbp]
\includegraphics[width=80mm]{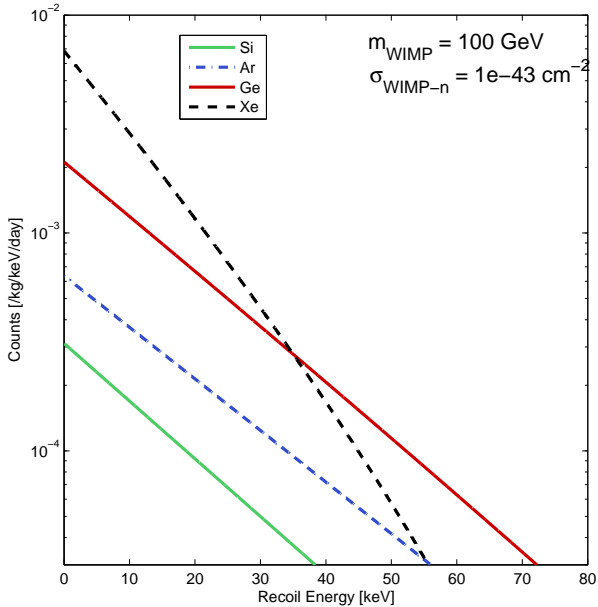}
\caption{Differential WIMP recoil spectrum for a WIMP mass of 100\,GeV 
and a WIMP-nucleon cross section $\sigma = 10^{-43}$ cm$^2$. The spectrum was 
calculated for illustrative nuclei such as Si (light solid), Ar (light dot-dashed), Ge (dark solid), Xe (dark dashed).} 
\label{rates}
\end{figure}

The nuclear form factor becomes significant at large WIMP and nucleus 
masses, and leads to a suppression of the differential scattering rate. 
Figure \ref{rates} shows differential spectra for Si, Ar, Ge and Xe, 
calculated for a WIMP mass of 100\,Gev, a WIMP-nucleon cross section of 
$\sigma = 10^{-43}$ cm$^2$ and using the standard halo parameters 
mentioned above.

A WIMP with a typical mass between a few GeV and 1\,TeV 
will deposit a recoil energy below 50\,keV in a terrestrial detector.
As for the predicted event rates for neutralinos, scans of the MSSM parameter 
space under additional assumptions  (GUT, mSUGRA, etc) and accounting for
accelerator and cosmological constraints, yield  about 10$^{-6}$ to 10
events per kilogram detector material and day \cite{theo_rates}.

Evidently, in order to observe a WIMP spectrum, 
low energy threshold, low background and high mass detectors are essential.
In such a detector, the recoil energy of the scattered nucleus is transformed into
a measurable signal, such as charge, light or phonons, 
and at least one of the above quantities is detected. Observing two  
signals simultaneously yields a powerful discrimination against background events,  
which are mostly interactions with electrons, as opposed to WIMPs and neutrons scattering 
off nuclei (see Section \ref{experiments} for a more detailed discussion). 
Even for experiments with good event by event discrimination, an absolute low background   
is still important. It can be achieved in both passive and active ways. Passive methods 
range from high material selection of detector
components to various specific shieldings against the natural radioactivity of the environment 
and against cosmic rays and secondary particles produced in their interactions.
Active background reduction implies an active shield, commonly a plastic or liquid 
scintillator surrounding the detector. An additional advantage is provided by a highly granular 
detector (or good timing and position resolution), 
since multiple scatters within the detector volume allow both background reduction 
and a direct measurement of the neutron background for experiments with an event-by-event 
discrimination against electron recoils.

In order to convincingly detect a WIMP signal, a specific signature from a particle populating  
our galactic halo is important.
The Earth's motion through the galaxy induces both a seasonal variation of 
the total event rate \cite{fre86,fre88} and a forward-backward 
asymmetry in a directional signal \cite{spergel88,copi99}. 

The annual modulation of the WIMP signal arises because of the Earth's motion 
in the galactic frame, which is a superposition of the Earth's rotation
around the Sun and the Sun's rotation around the galactic center:

\begin{equation}
v_E = v_\odot + v_{\rm orb} \; \cos \gamma \; \cos \omega (t - t_0), 
\end{equation}

where $v_\odot = v_0$ + 12 km s$^{-1}$ ($v_0$ $\approx$ 220  km s$^{-1}$),  
$v_{\rm orb}$ $\approx$ 30 km s$^{-1}$ denotes 
the Earth's orbital speed around the Sun, the angle 
$\gamma \approx 60^0$ is the inclination of the Earth's orbital plane 
with respect to the galactic plane and $\omega = 2 \pi / 1 \mbox{yr}$, 
$t_0 =$ June  2$^{\rm nd}$.  

The expected time dependence of the count rate can  be approximated 
by a cosine function with a period of T = 1\,year and a phase of $t_0
=$ June  2$^{\rm nd}$:

\begin{equation}
S(t) = S_0 + S_m cos \omega (t - t_0),
\end{equation}

where $S_0$, $S_m$ are the constant and the modulated amplitude of the signal, 
respectively.
In reality, an additional contribution to S(t) from the
background must be considered.  The background should  be constant in time or at least 
not show the same time dependency as the predicted WIMP signal.

The expected seasonal modulation effect is very small 
(of the order of $v_{\rm orb}$/$v_0 \simeq $ 0.07), requiring
large masses and long counting times as well as an excellent long-term stability 
of the experiment.

A much stronger signature would be given by the ability to detect the axis and 
direction of the recoil nucleus.
In \cite{spergel88} it has been shown, that the WIMP interaction rate as a function 
of recoil energy and angle $\theta$ between the WIMP velocity and
recoil direction 
(in the galactic frame) is:

\begin{equation}
\frac{d^2R}{dE_R d cos\theta} \propto \rm{exp}\left[-\frac{(v_\odot cos\theta - v_{min})^2}{v_0^2} \right],  
\end{equation}

where v$_{min}^2$ = (m$_N$ + m$_W$)$^2$E$_R$/2m$_N$m$_W^2$ 
and v$_0^2$=3v$_\odot^2$/2.

The forward-backward asymmetry thus yields a large effect of the order of 
$\mathcal{O}$(v$_\odot$/v$_0$)$\approx$ 1 and fewer events are needed 
to discover a WIMP signal than in the case of the seasonal modulation \cite{copi99}.
The challenge is to build massive detectors capable of detecting the direction of the 
incoming WIMP. At present, only one such detector exists (DRIFT \cite{drift}) with 
a total active mass of $\sim$170\,g  of CS$_2$, with larger modules 
being under consideration.

\section{\label{experiments}Experiments}

First limits on WIMP-nucleon cross sections were derived about twenty years ago, 
from at that time already existing germanium double beta decay experiments \cite{ge_exp}. 
With low intrinsic backgrounds and already operating in underground 
laboratories, these detectors were essential in ruling out first WIMP candidates such as 
a heavy Dirac neutrino \cite{beck94}. Present Ge ionization experiments dedicated to 
dark matter searches such as HDMS \cite{hdms} are limited in their sensitivity by 
irreducible electromagnetic backgrounds close to the crystals or from 
cosmogenic activations of the crystal themselves.
Next generation projects based on high-purity germanium (HPGe) ionization detectors, 
such as the proposed GENIUS \cite{genius}, GERDA \cite{gerda}, and Majorana \cite{majorana} experiments, 
aim at an absolute background reduction by more than three orders in magnitude, compensating for 
their inability to differentiate between electron- and nuclear recoils on an event-by-event basis. 
Note that the primary scientific goal of these projects is to look for the neutrinoless 
double beta decay, at a Q-value around 2039\,keV in $^{76}$Ge.
Solid scintillators operated at room temperatures had soon caught up with HPGe experiments, 
despite their higher radioactive backgrounds. Being intrinsically fast, these experiments 
can discern on a statistical basis between electron and nuclear recoils, by using the timing 
parameters of the pulse shape of a signal. Typical examples are NaI experiments 
such as DAMA \cite{dama} and NAIAD \cite{naiad}, with DAMA reporting first evidence for a positive 
WIMP signal in 1997 \cite{dama-wimp}. The DAMA results have not been confirmed by three different 
mK cryogenic experiments (CDMS \cite{cdms,cdms-04}, CRESST \cite{cresst04} and EDELWEISS \cite{edelweiss,edelw05}) and 
one liquid xenon experiment (ZEPLIN \cite{bdmc-04}), independent of the halo model assumed \cite{halo-dama} 
or whether the WIMP-nucleon interaction is taken as purely spin-dependent \cite{spin-dama,savage04}. 
The DAMA collaboration has installed a new, 250\,kg NaI experiment (LIBRA) in the Gran Sasso Laboratory,   
and began taking data in March 2003. With lower backgrounds and increased statistics,  
LIBRA should soon be able to confirm the annual modulation signal. 
The Zaragosa group plans to operate a 107\,kg NaI array (ANAIS) at the Canfranc 
Underground Laboratory (2450 mwe) in Spain \cite{zaragosa-NaI}, and deliver an independent check of the 
DAMA signal in NaI. 
Cryogenic experiments operated at sub-Kelvin temperatures are now leading the field with sensitivities 
of one order of magnitude above the best solid scintillator experiments. Specifically, 
the CDMS experiment can probe WIMP-nucleon cross sections as low as 10$^{-43}$cm$^2$ \cite{cdms-04}. 
This class of experiments will be covered in more detail in Section \ref{cryo_mK}. 
Liquid noble element detectors are rapidly evolving, and seem a very promising avenue towards 
the goal of constructing ton-scale or even multi-ton WIMP detectors. These approaches 
will be discussed in Section \ref{liquid_noble}. 
Many other interesting WIMP search techniques have been deployed, and it is not the scope of this paper 
to deliver a full overwiev. For a significantly more detailed accounting of existing and future detection 
techniques, as well as critical discussions of backgrounds and of the reported positive signal, we 
refer to two recent reviews \cite{gabriel05,rick04}.

\subsection{\label{cryo_mK}Cryogenic Detectors at mK Temperatures}

Cryogenic calorimeters are meeting crucial characteristics of a successful WIMP 
detector: low energy threshold ($<$10\,keV), excellent energy resolution ($<$1\% at 10\,keV)  
and the ability to differentiate nuclear from electron recoils on an event-by-event basis. 
Their development was driven by the exciting possibility of doing a calorimetric energy 
measurement down to very low energies with unsurpassed energy resolution. 
Because of the T$^3$ dependence of the heat capacity of a dielectric crystal, 
at low temperatures a small energy deposition can significantly change the temperature 
of the absorber. The change in temperature can be measured either after the phonons 
(or lattice vibration quanta) reach equilibrium, or thermalize, or when they are still 
out of equilibrium, or athermal, the latter providing additional information 
about the location of an event. The astounding energy resolution comes from the fact that 
a much lower energy ($<$1\,meV) is required to produce an elementary phonon excitation 
compared to semiconductor detectors ($\sim$1\,eV). If a second signal, such as charge or 
scintillation light is detected, identification of background events induced by electron 
recoils is possible. It is based on the measured difference in the ratio of charge or 
light to phonons for electron and nuclear recoils. If a class of events (such as, for example, 
interactions within tens of micrometers of the detector's surface) results in incomplete charge 
or light collection, then such events can leak into the parameter region predicted for nuclear 
recoils. As will be discussed later, no such events have been observed yet in phonon-light detectors. 
On the other hand, the timing properties of the fast phonon signal in certain phonon-charge 
detectors allow a discrimination between surface events and interactions in the bulk.

\subsubsection{CDMS}

\begin{figure}[thbp]
\includegraphics[width=80mm]{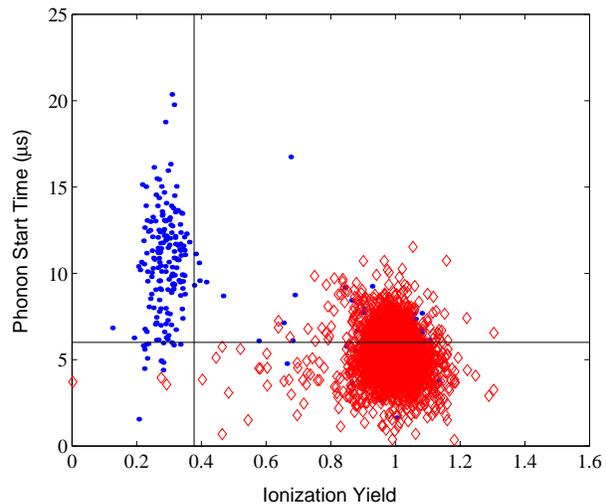}
\caption{Phonon start times versus ionization yield for $^{133}$Ba gamma calibration events (diamonds)
and $^{252}$Cf neutron calibration events (dots). The grey lines indicate timing and ionization-yield 
cuts, resulting in a high-rate of nuclear recoil efficiency and a low rate of misidentified surface events.} 
\label{starttimes}
\end{figure}

The Cold Dark Matter Search experiment operates low-temperature Ge and Si detectors at the 
Soudan Underground Laboratory in Minnesota (at a depth of 2080 m.w.e.).  The high-purity 
Ge and Si crystals are 1\,cm thick and 7.6\,cm in diameter, and have a mass of 250\,g and 
100\,g, respectively. They are operated at a base temperature around 50\,mK. Superconducting 
transition edge sensors photolitographically patterned onto one of the crystal surfaces 
detect the athermal phonons from particle interactions. The phonon sensors are divided into 
4 different channels, allowing to reconstruct the x-y position of an event with a resolution 
of $\sim$1\,mm. If an event occurs close to the detector's surface, the phonon signal is faster 
than for events far from the surface, because of phonon interactions in the thin metallic films.
The risetime of the phonon pulses, as well as the time difference between the charge and phonon 
signals allow to reject surface events caused by electron recoils. Figure~\ref{starttimes} 
shows phonon start times versus ionization yield (charge energy divided by the total recoil energy) 
for electron recoil events (collected with a $^{133}$Ba source) and nuclear recoil events 
(collected with a $^{252}$Cf source). Events below a yield around 0.75 typically occur within 
0-30\,$\mu$m of the surface, and can be effectively discriminated 
(a typical rise time cut is shown by the horizontal grey line in the figure) 
while preserving a large part of the nuclear recoil signal.

\begin{figure}[thbp]
\includegraphics[width=80mm]{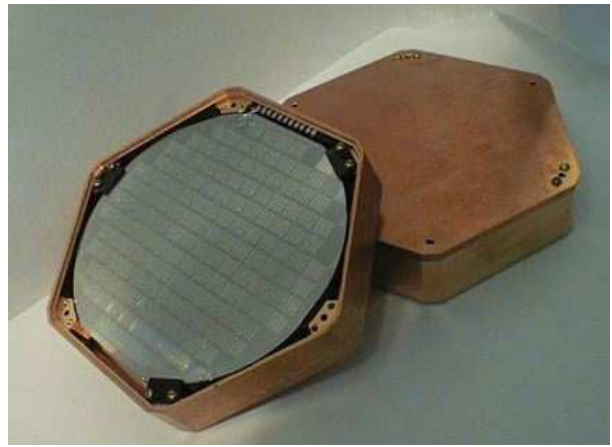}
\caption{CDMS detectors (250\,g Ge or 100\,g Si, 1\,cm thick and 7.5\,cm diameter) in their Cu holders.} 
\label{zips}
\end{figure}

Charge electrodes are used for the ionization measurement. They are divided into an inner disk, 
covering 85\% of the surface, and an outer guard ring, which is used to reject events near the 
edges of the crystal, where background interactions are more likely to occur. 
Figure \ref{zips} shows a picture of CDMS detectors in their Cu holders.

The discrimination against the electron recoil background is based on the fact that 
nuclear recoils (caused by WIMPs or neutrons) produce fewer charge pairs than electron 
recoils of the same energy. The ionization yield, defined as the ratio of ionization to recoil 
energy, is about 0.3 in Ge, and 0.25 in Si for recoil energies above 20\,keV. Electron 
recoils with complete charge collection show an ionization yield of $\approx$1.  
For recoil energies above 10\,keV, bulk electron recoils are rejected with $>$99.9\% efficiency,
and surface events are rejected with $>$95\% efficiency.
The two different materials are used to distinguish between WIMP and neutron interactions by 
comparing the rate and the spectrum shape of nuclear recoil events. While the interaction rate 
of neutrons is comparable in Ge and Si, the WIMP rate is much higher in Ge for spin-independent 
couplings. 

\begin{figure}[thbp]
\centering
\includegraphics[width=80mm]{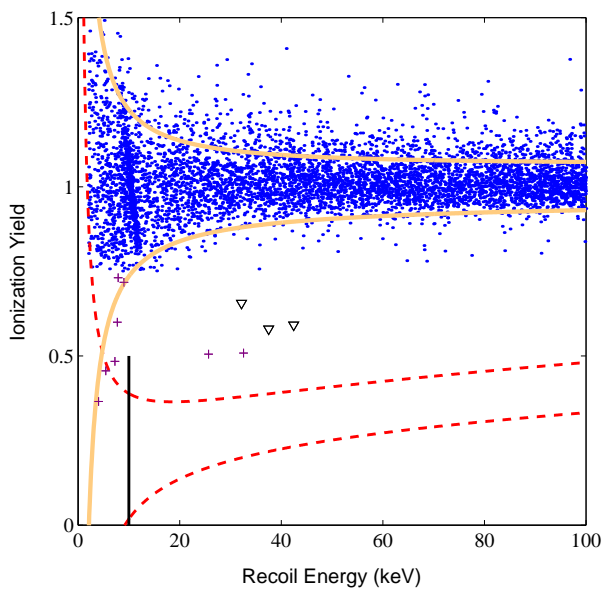}
\caption{Ionization yield versus recoils energy for 3 Ge detectors of Tower\,1 for a total lifetime 
of 53\,kg\,days. Events above an ionization yield of 0.75 are shown as points (note the 10.4\,keV Ga X-ray line), 
events below a yield of 0.75 are shown as geometrical shapes. The expected signal region for WIMP recoils 
lies between the two dashed lines around a yield of 0.3. The dark vertical line denotes the analysis threshold 
of 10\,keV recoil energy.} 
\label{cdms_y_erecoil}
\end{figure}

A stack of six Ge or Si detectors together with the corresponding cold electronics is named a 'tower'.
Five towers (30 detectors) are currently installed in the 'cold volume' at Soudan, shielded by about 3\,mm of Cu, 
22.5\,cm of Pb, 50\,cm of polyethylene and by a 5\,cm thick plastic scintillator detector  
which identifies interactions caused by cosmic rays penetrating the Soudan rock.
Results from the first tower of 4 Ge and 2 Si detectors \cite{cdms-04}, which took data at Soudan 
from October 2003 to January 2004, showed no nuclear recoil candidate in 53 raw live days 
(see Figure \ref{cdms_y_erecoil}).

\begin{figure}[thbp]
\centering
\includegraphics[width=80mm]{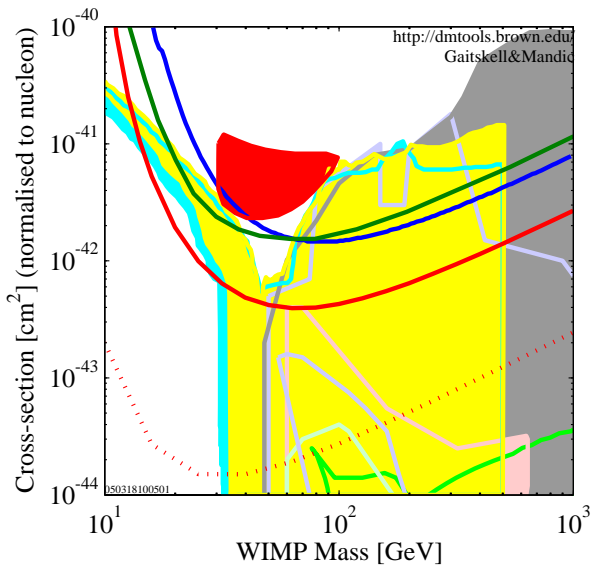}
\caption{Experimental results and theoretical predictions for spin-independent WIMP nucleon cross sections 
versus WIMP mass. The data (from high to low cross sections) show the DAMA allowed region (red) \cite{dama-wimp}, 
the latest EDELWEISS result (blue) \cite{edelw05}, the ZEPLIN\,I preliminary results (green) \cite{bdmc-04} 
and the CDMS results from 
Tower\,1 at Soudan (red) \cite{cdms-04}. Also shown is the expectation for 5 CDMS towers at Soudan (red dashed). 
The SUSY theory regions are shown as filled regions or contour lines, and are taken from 
\cite{susy-theory}.}
\label{limits}
\end{figure}

 After cuts, the net exposure was 22\,kg\,days for Ge and 5\,kg\,days for Si.
The resulting upper limit on WIMP-nucleon cross sections for spin-independent couplings and a standard halo 
is 4$\times$10$^{-43}$cm$^2$ at the 90\% C.L. at a WIMP mass of 60\,GeV (see Figure~\ref{limits}), 
which is four times below the best previous limit reported by EDELWEISS \cite{edelweiss}.

\begin{figure}[thbp]
\centering
\includegraphics[width=80mm]{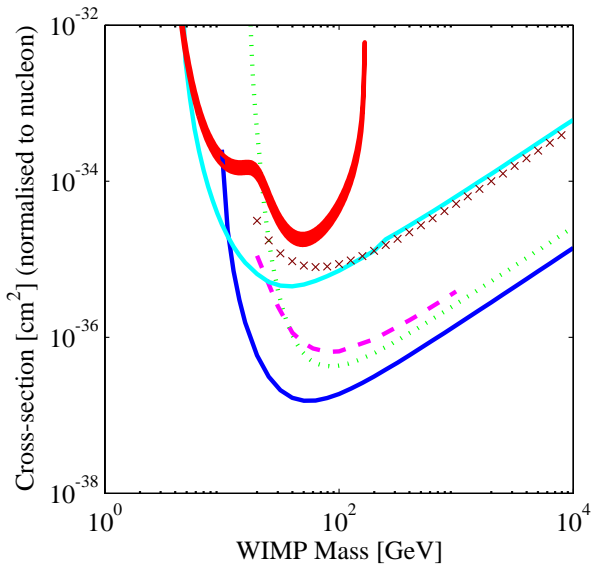}
\caption{Experimental results for spin-dependent WIMP couplings (90\% C.L. contours), 
for the case of a pure neutron coupling. The curves (from high to low cross sections) 
show the DAMA annual modulation signal (filled red region), the CDMS Soudan Si data 
(red crosses), the CDMS Stanford Si data (cyan),  EDELWEISS (magenta dashed), 
DAMA/Xe (green dotted) and the CDMS Soudan Ge data (solid blue). For details 
and references see \cite{jeff05}.} 
\label{limits-cdms-sd}
\end{figure}

The limits on spin-dependent WIMP interactions are competitive with other experiments, 
in spite of the low abundance of $^{73}$Ge (7.8\%) in natural germanium. In particular, in the case 
of a pure neutron coupling, CDMS yields the most stringent limit obtained so far, 
thus strongly constraining interpretations of the DAMA signal region \cite{savage04,cdms-sd} (Figure~\ref{limits-cdms-sd}). 
For details on the CDMS analysis for spin-dependent WIMP couplings we refer to 
Jeff Filippini's contribution to these proceedings \cite{jeff05}.

Two towers (6 Ge and 6 Si detectors) were operated at Soudan from March to August 2004.
The data is being currently analyzed in a similar fashion to the first Soudan run, in which all cuts 
are determined in a blind manner from calibration data, without access to the signal region.
Calibration runs were interspersed with science runs, and taken with a $^{252}$Cf (neutron and gamma) 
and a $^{133}$Ba (gamma) source. The high-statistics Ba data also allows for a calibration to surface events, 
providing a population of so called 'ejectrons', or low energy electrons which are ejected from an adjacent 
detector of from material close to a detector's surface.
For an overview of the currently running experiment at Soudan and expectations from the 5\,tower run in 2005,
we refer to Walter Ogburn's contribution to these proceedings \cite{walter05}.

\subsubsection{EDELWEISS}

\begin{figure}[thbp]
\includegraphics[width=80mm]{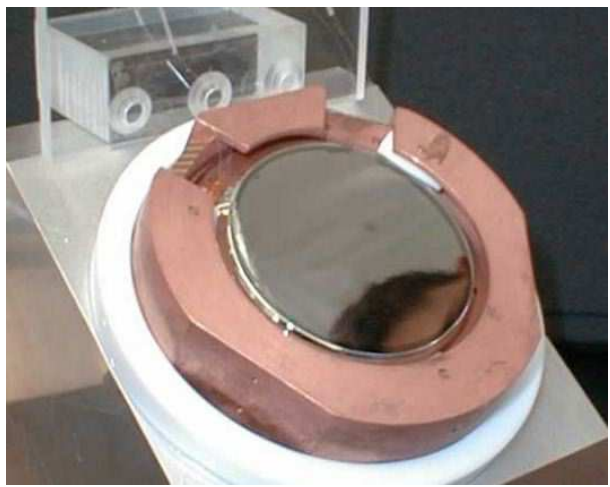}
\caption{Photograph of a 320\,g EDELWEISS germanium bolometer.} 
\label{edelw-det}
\end{figure}

The EDELWEISS experiment operates germanium bolometers at 17\,mK in the Laboratoire 
Souterrain de Modane, at about 4800\,m.w.e. The detectors are further shielded by 30\,cm 
of paraffin, 15\,cm of Pb and 10\,cm of Cu. As in the case of CDMS, they simultaneously detect 
the phonon and the ionization signals, allowing a discrimination against bulk electron 
recoils of better than 99.9\% above 15\,keV recoil energy. The charge signal is measured by Al 
electrodes sputtered on each side of the crystals, the phonon signal by a neutron transmutation 
doped (NTD) heat sensor glued onto one of the charge collection electrodes. The NTD sensors read out 
the thermal phonon signal on a time scale of about 100\,ms. Figure~\ref{edelw-det} shows a picture 
of a 320\,g germanium bolometer.

\begin{figure}[thbp]
\centering
\includegraphics[width=80mm]{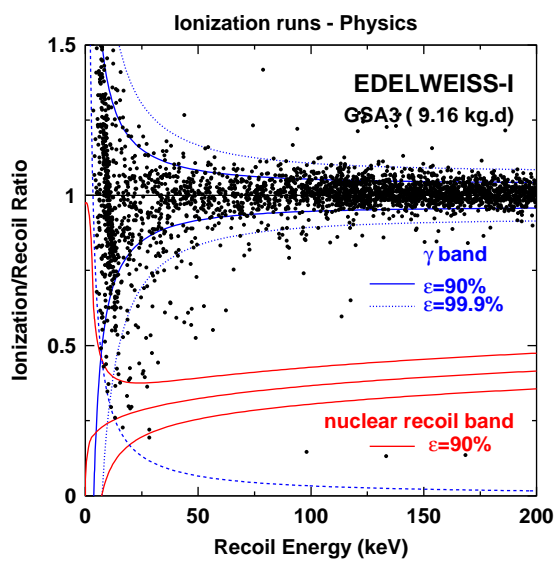}
\caption{Ionization yield versus recoil energy for one EDELWEISS 320\,g Ge detector with an 
exposure of 9.16\,kg\,days. Also shown are the electron recoil (blue) and neutron recoil (red) bands.
Figure taken from \cite{edelw05}.} 
\label{edelw-yplot}
\end{figure}

Between 2000-2003, EDELWEISS performed four physics runs with five 320\,g Ge crystals, accumulating 
a total exposure of 62\,kg\,days \cite{sanglard04}. Above an analysis threshold of 20\,keV, 
a total of 23 events compatible with nuclear recoils have been observed. Figure \ref{edelw-yplot}
shows the ionization yield versus recoil energy for one EDELWEISS detector for an exposure of 
9.16\,kg\,days. 

\begin{figure}[thbp]
\centering
\includegraphics[width=80mm]{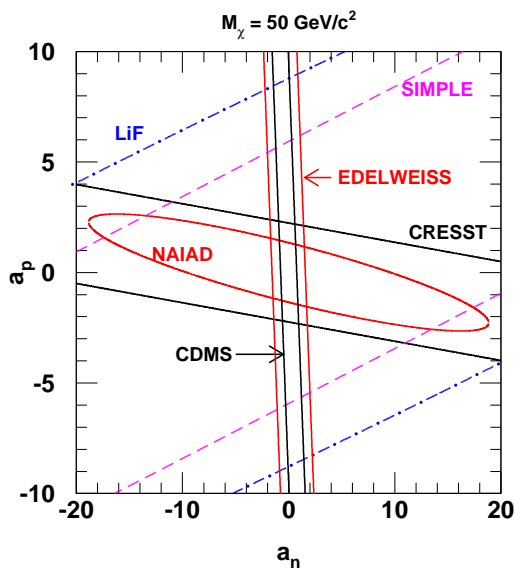}
\caption{Allowed regions in the ($a_n$,$a_p$) plane for a WIMP mass of 50\,GeV. $a_{p,n}$ are the effective WIMP 
couplings to proton and neutrons (see \cite{edelweiss-sd} for details). 
The allowed regions are those within the parallel straight lines or ellipses.} 
\label{limits-sd}
\end{figure}

The derived upper limit 
on spin-independent WIMP-nucleon couplings under the hypothesis that all above events are caused by WIMP 
interactions, and for a standard isothermal halo, is shown in Figure~\ref{limits}.
The allowed region in the ($a_n$,$a_p$) plane for spin-dependent WIMP couplings is shown in 
Figure~\ref{limits-sd}, for a
WIMP mass of 50\,GeV ($a_{p,n}$ are the effective WIMP 
couplings to proton and neutrons). 

The EDELWEISS experiment has ceased running in March 2004, in order to allow the upgrade to a second phase, 
with an aimed sensitivity of 10$^{-44}$cm$^2$. The new 50 liter low-radioactivity cryostat will be able to house 
up to 120 detectors. Because of the inability of slow thermal detectors 
to distinguish between low-yield surface events and nuclear 
recoils and the inherent radioactivity of NTD sensors, the collaboration has been developing a new design 
based on NbSi thin-film sensors. These films, besides providing a lower mass and radioactivity per sensor, 
show a strong difference in the pulse shape, depending on the interaction depth of an event \cite{nader01}.
The EDELWEISS collaboration plans to operate twenty-one 320\,g Ge detectors equipped with NTD sensors, 
and seven 400\,g Ge detectors with NbSi thin-films in the new cryostat starting in 2005.

\subsubsection{CRESST}

\begin{figure}[thbp]
\includegraphics[width=80mm]{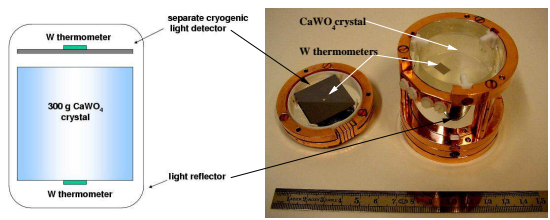}
\caption{Schematic picture and photograph of a CRESST CaWO$_4$ light and phonon detector. 
Figure taken from \cite{cresst04}.} 
\label{cresst-det}
\end{figure}

The CRESST collaboration has developed cryogenic detectors based on CaWO$_4$ crystals, 
which show a higher light yield at low temperatures compared to other scintillating materials.  
The detectors are also equipped with a separate, cryogenic light detector made of a 30$\times$30$\times$0.4\,mm$^3$
silicon wafer, which is mounted close to a flat surface of the CaWO$_4$ crystal. 
The temperature rise in both CaWO$_4$ and light detector is measured with  tungsten superconducting 
phase transition thermometers, kept around 10\,mK, in the middle of their transition between the superconducting 
and normal conducting state. A schematic picture and a photograph of a CRESST detector is shown 
in Figure~\ref{cresst-det}. 

A nuclear recoil in the 300\,g CaWO$_4$ detector 
has a different scintillation light yield than an electron recoil of the same energy, allowing to 
discriminate between the two type of events when both the phonon and the light signals are observed. 
The advantage of CaWO$_4$ detectors is their low energy threshold in the phonon signal, and the fact 
that no light yield degradation for surface events has been detected so far. 
However, about 1\% or less of the energy deposited in the CaWO$_4$ is seen as scintillation light 
\cite{cresst04}. Only a few tens of photons are emitted per keV electron recoil, a number which is further 
diminished for nuclear recoils, because of the involved quenching factor. 
The quenching factor of oxygen nuclear recoils for scintillation light is around 
13.5\% relative to electron recoils \cite{cresst04}, leading to a rather high effective recoil energy threshold for  
the detection of the light signal. While neutrons will scatter predominantly on oxygen nuclei, 
it is expected that WIMPs will more likely scatter on the heavier calcium and tungsten. The quenching factor of tungsten 
at room temperatures has been measured to be around 2.5\% \cite{cresst04}, making it difficult 
to observe the light signal of WIMP recoils above the thermal noise.

The most recent CRESST results were obtained by operating two 300\,g CaWO$_4$ detectors at the Gran Sasso 
Underground Laboratory (3800 m.w.e) for two months at the beginning of 2004 \cite{cresst04}. 
The total exposure after cuts  was 20.5\,kg\,days. The energy resolution in the phonon channel was 
 1.0\,keV (FWHM) at 46.5\,keV, the low-energy gamma line being provided by an external $^{210}$Pb contamination.  
A total of 16 events were observed in the 12\,keV - 40\,keV recoil energy region, 
a number which seems consistent with the expected neutron background, since 
the experiment had no neutron shield at this stage. 
No phonon-only events (as expected for WIMP recoils on tungsten) were observed between 12\,keV - 40\,keV 
in the module with better resolution in the light channel, yielding a limit on coherent WIMP 
interaction cross sections very similar to the one obtained by EDELWEISS.

CRESST has stopped taking data in March 2004, to upgrade with a neutron shield, an active muon veto, 
and a 66-channels SQUID read-out system. It will allow to operate 33 CaWO$_4$ detector modules, providing 
a total of 10\,kg of target material. The upgrade will be completed in early 2005 and the expected sensitivity 
is around 10$^{-44}$cm$^2$.

\subsection{\label{liquid_noble}Liquid Xenon Detectors }

Liquid xenon (LXe) has excellent properties as a dark matter detector. It has a high density (3\,g/cm$^3$) 
and high atomic number (Z=54, A=131.3), allowing experiments to be compact. The high mass of the xenon nucleus 
is favorable for WIMP scalar interactions provided that a low energy threshold can be achieved (see Figure~\ref{rates} 
for a comparison with other target nuclei). LXe is an intrinsic scintillator, having high scintillation ($\lambda$ = 178~nm) 
and ionization yields because of its low ionization potential (12.13~eV).
There are no long-lived radioactive xenon isotopes, and other impurities (such as $^{85}$Kr) can be reduced to a very low 
level by centrifugation or with a distillation tower and a cold trap. Krypton contamination levels as low as 1\,ppb 
(parts per billion) have already been achieved \cite{xenon_proposal}.

Scintillation in liquid xenon is produced by the formation of excimer states, which are 
bound states of ion-atom systems. If a high electric field ($\sim$1 kV/cm) is applied, ionization electrons can also
be detected, either directly or through the secondary process of proportional scintillation.

The elastic scattering of a WIMP produces a low-energy xenon recoil,  which loses its energy through ionization and scintillation. 
Both signals are quenched when compared to an electron recoil of the same energy, but by different amounts, allowing 
to use the ratio for distinguishing between electron and nuclear recoils. The quenching factor for both scintillation 
and ionization depends on the drift field and on the energy of the recoil. 

At zero electric field, the relative scintillation efficiency of nuclear
recoils in LXe was recently measured to be in the range of 0.13-0.23 for Xe
recoil energies of 10 keV-56 keV \cite{elenaprd-05}. 

There are three efforts to develop liquid xenon detectors operated in the dual-phase (liquid and gas) mode. 

\subsubsection{ZEPLIN}

\begin{figure}[thbp]
\includegraphics[width=80mm]{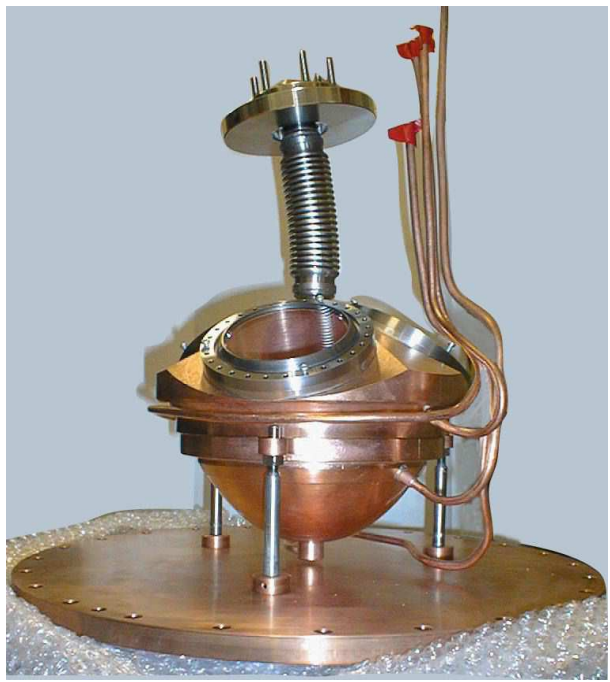}
\caption{Photograph of the ZEPLIN\,I inner liquid xenon detector.}
\label{zeplin-det}
\end{figure}

The Boulby Dark Matter collaboration has been operating a single-phase liquid xenon detector, ZEPLIN\,I, at the Boulby Mine 
($\sim$3000 m.w.e.) during 2001-2002. The ZEPLIN\,I detector had a fiducial mass of 3.2\,kg of liquid xenon, 
viewed by 3 PMTs through silica windows. It was inclosed in a 0.93 ton active scintillator veto, 
which helped in reducing the background from the radioactivity of the PMTs and surroundings. 
A picture inner detector is shown in Figure~\ref{zeplin-det}.

A total exposure of 293\,kg\,days had been accumulated. With a light yield of 1.5 electrons/keV, the energy 
threshold was at 2\,keV electron recoil (corresponding to 10\,keV nuclear recoil energy for a 
quenching factor of 20\%). A discrimination between electron and nuclear recoils was applied by using  
the difference in the mean time of the corresponding pulses. To establish this difference, the detector had 
been calibrated with gamma sources (above ground and at the mine), and with neutron sources above ground.
Using this statistical discrimination, 
a preliminary limit on spin-independent WIMP cross sections comparable to CRESST and EDELWEISS has been achieved 
(see Figure~\ref{limits}).

The collaboration has developed two concepts for dual-phase detectors, ZEPLIN\,II and ZEPLIN\,III.
ZEPLIN\,II will have a 30\,kg fiducial target mass, observed by 7 PMTs. 
ZEPLIN\,III will operate a lower target mass (6\,kg liquid xenon viewed by 31 PMTs)  at a higher field ($>$ 5\,kV/cm). 
It is expected that both ZEPLIN\,II and ZEPLIN\,III will be deployed at the Boulby Mine in 2005.

\subsubsection{XMASS}

The Japanese XMASS collaboration develops liquid xenon detectors for solar neutrinos, double beta decay and dark matter 
searches \cite{xmass-04}. Since 2003, they have been operating a 1\,kg dual-phase detector at the Kamioka Mine (2700 mwe).
The active volume was viewed by two UV-sensitive PMTs through MgF$_2$ windows, the detector being operated 
in the low drift field regime (250\,V/cm) \cite{masaki-03}. At such fields, the proportional light signal is large for
electron recoils, but very small or zero for nuclear recoils. Events which yield a signal only in the light channel 
can thus mimic a potential WIMP candidate. A second prototype with an active volume of 15\,kg viewed by 14 PMTs 
is currently under operation at the Kamioka Underground Laboratory. A 100\,kg single-phase detector is also being operated 
at Kamioka \cite{xmass-04}. Thirty liters of liquid xenon are contained in a (31\,cm)$^3$ oxygen free copper vessel, 
and viewed by 54 low-background, 2 inch PMTs. So far, the background levels are consistent within a factor of 
two with expectations, although an internal $^{222}$Rn source has been detected \cite{xmass-04}.

\subsubsection{XENON}

\begin{figure}[thbp]
\centering
\includegraphics[width=80mm]{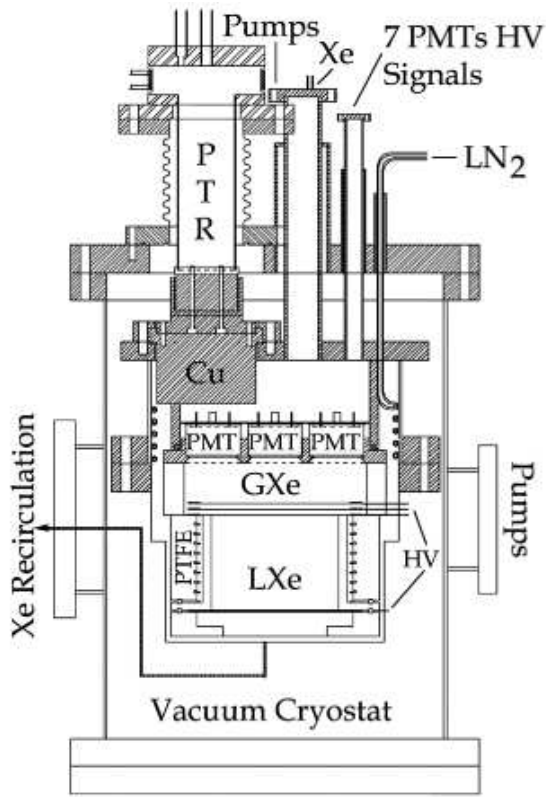}
\caption{Schematic drawing of the 3\,kg XENON dual phase prototype.} 
\label{3kg-schematic}
\end{figure}

\begin{figure}[thbp]
\centering
\includegraphics[width=80mm]{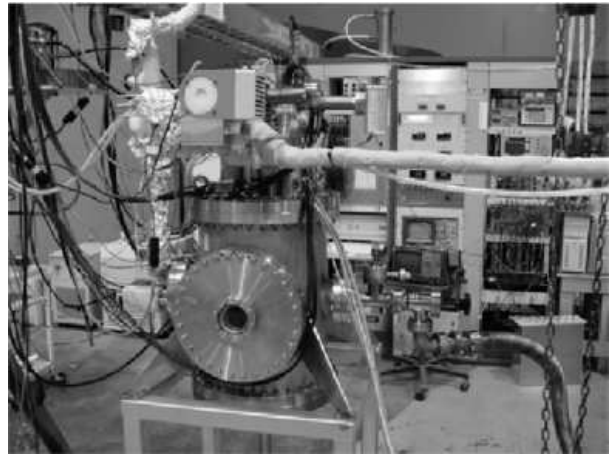}
\caption{Picture of the 3\,kg (active mass) XENON dual phase cryostat.} 
\label{3kg-picture}
\end{figure}

The US XENON collaboration plans to deploy and operate a 10\,kg dual-phase detector in the Gran Sasso Underground 
Laboratory (3500 m.w.e) by 2005-2006. At present, a 3\,kg prototype is under operation above ground, 
at the Columbia Nevis Laboratory \cite{elena-05}. The detector is operated at a drift field of 1\,kV/cm, and
both primary and proportional light are detected by an array of seven 2\,inch PMTs operating in 
the cold gas above the liquid. The active volume is defined by PTFE walls and Be-Cu wire grids with high optical transmission.
The detector is insulated by a vacuum cryostat and cooled down to a stable temperature of (-100$\pm$0.05)\,C with a pulse tube 
refrigerator (see Figure~\ref{3kg-schematic} for a schematic drawing of the prototype and Figure~\ref{3kg-picture} 
for a picture of the cryostat.). 

The xenon gas is continuously 
circulated and purified using a high temperature SAES getter \cite{saes}.
The performance of the chamber was tested with gamma ($^{57}$Co), alpha ($^{210}$Pb) and neutron ($^{241}$AmBe) sources.
The depth of an event is reconstructed by looking at the separation in time between the primary and proportional 
scintillation signal (the electron drift velocity for the applied electric field is known). The x-y position is inferred 
with a resolution of 1\,cm from the center of gravity of the proportional light emitted close to the seven PMTs.
The measured ratio of proportional light (S2)  to direct light (S1) for alpha recoils is 0.03 if the corresponding 
ratio for gamma events (electron recoils) is normalized to 1, providing a very clear separation between these type of events. 

\begin{figure}[thbp]
\centering
\includegraphics[width=80mm]{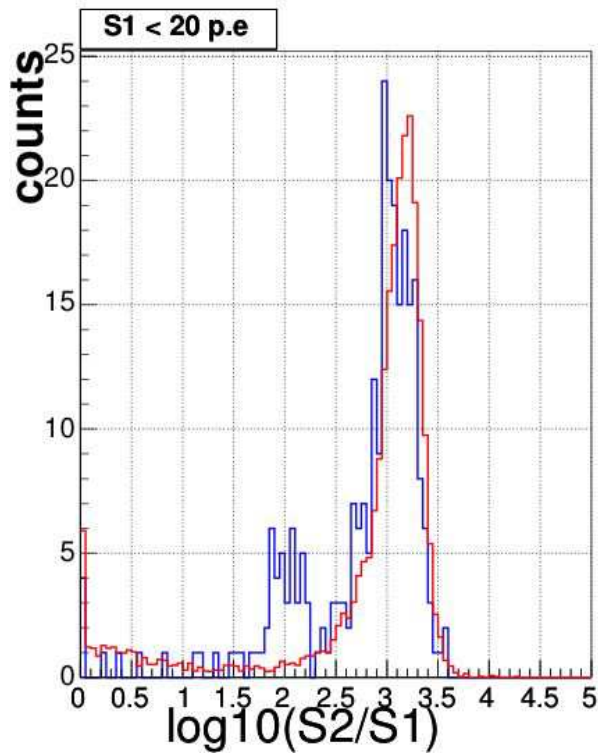}
\caption{Histogram showing the S2/S1 distribution for AmBe events (blue) versus 662\,keV gamma events 
from $^{137}$Cs (red). Two distinct populations are visible in the AmBe data. Figure taken from \cite{elena-05}.} 
\label{xenon-nrecoils}
\end{figure}

More interesting is the ratio S2/S1 for nuclear recoil events. It was established using a $^{241}$AmBe neutron source, 
by selecting events which were tagged as neutron recoils in a separate neutron detector placed under a scattering 
angle of 130\,deg. If the S2/S1 ratio for electron recoils (provided by a $^{137}$Cs source) is again 
normalized to 1, then S2/S1 for nuclear recoils was measured to be around 0.1, the leakage of electron recoils into the 
S2/S1 region for nuclear recoils being $<1\%$ (for details see \cite{elena-05}). Figure \ref{xenon-nrecoils} 
shows a histogram of S2/S1 for events taken with the $^{241}$AmBe source, compared to the corresponding distribution 
of events from the $^{137}$Cs gamma source.

These first measurements 
are very encouraging, and will be further improved by using a CsI photocathode immersed in the liquid. The 
downwards going primary photons are converted into electrons, which are then drifted into the gas phase, 
where electroluminiscence occurs and a tertiary signal is observed by the PMTs. First tests of a CsI 
photocathode in the 3\,kg prototype have confirmed the high quantum efficiency (around 12\% at 1\,kV/cm) which 
had been previously measured \cite{elena-94}.

The first XENON detector with a fiducial mass of 10\,kg (XENON10) to be operated in Gran Sasso is currently 
under construction. Its goal is to achieve a sensitivity of a factor of 20 below the current CDMS results, 
thus probing WIMP cross sections around 2$\times$10$^{-44}$cm$^2$.

\subsection{The Future}

We live in suspenseful times for the field of direct detection: for the first time, a couple of experiments 
operating deep underground probe the most optimistic supersymmetric models. The best limits on WIMP-nucleon 
cross sections come from cryogenic experiments with ultra-low backgrounds and excellent event-by-event 
discrimination power (CDMS, EDELWEISS, CRESST). Although these experiments had started with target masses  
around 1\,kg, upgrades to several kilograms have already taken place or are foreseen for the near future, 
ensuring (along with improved backgrounds) an increase in sensitivity by a factor of 10-100. Other 
techniques, using liquid noble elements such as xenon, may soon catch up and probe similar parameter spaces 
to low-temperature cryogenic detectors. It is worth emphasizing here that given the importance 
of the endeavor and the challenge in unequivocally identifying and measuring the 
properties of a dark matter particle, it is essential that more than one technique will move forward. 

If supersymmetry is the answer to open questions in both cosmology and particle physics, then 
WIMP-nucleon cross sections as low as 10$^{-48}$cm$^2$ are likely \cite{ellis05}. Likewise, in theories with universal 
extra dimensions, it is predicted that the lightest Kaluza Klein particle would have a scattering 
cross section with nucleons in the range of 10$^{-46}$ - 10$^{-48}$cm$^2$ \cite{servant-02}. 
Thus, to observe a signal of a few events per year, ton or even multi-ton experiments are inevitable. 
There are several proposals to build larger and improved dark matter detectors 
(see \cite{rick04} for an exhaustive list). The selection presented below is likely biased, but  based 
on technologies which seem the most promising to date. 

The SuperCDMS project is a three-phase proposal to 
utilize CDMS-style detectors with target masses growing  from 27\,kg to 145\,kg and up to 1100\,kg, with 
the aim of reaching a final sensitivity of  3$\times$10$^{-46}$cm$^2$ by mid 2015.  This goal 
will be realized by developing improved detectors (for a more precise event reconstruction) and 
analysis techniques, and at the same time by strongly reducing the intrinsic surface contamination 
of the crystals. A possible site for SuperCDMS is the recently approved SNO-Lab Deep-site facility 
in Canada (at 6000 m.w.e.), where the neutron background would be reduced by more than two orders 
of magnitude compared to the Soudan Mine, thus ensuring the mandatory conditions to build a
zero-background experiment. For details on the SuperCDMS project we refer to Paul Brink's contribution 
to these proceedings \cite{paul-05}. In Europe, a similar project to develop a 100\,kg-1\,ton 
cryogenic experiment, EURECA (European Underground Rare Event search with Calorimeter Array) 
\cite{gabriel05} is underway. 
The XENON collaboration is designing a 100\,kg scale dual-phase xenon detector (XENON100), towards 
a modular one tonne experiment. The baseline goal is 99.5\% background rejection efficiency above 
a 16\,keV nuclear recoil energy threshold, full 3-D localization of an event and a liquid xenon self shielding 
surrounding the detector. The aim of XENON100 is to probe the 
parameter space down to  cross sections of 10$^{-45}$cm$^2$. The full scale, one tonne experiment 
(XENON1t), which will operate ten XENON100 modules, will increase this sensitivity by another order 
of magnitude \cite{elena-05}.  ZEPLIN MAX, a R\&D project of the Boulby Dark Matter collaboration, 
is a further proposal to build a ton scale liquid xenon experiment. The design will be based on 
the experience and results with ZEPLIN II/III at the Boulby Mine \cite{bdmc-04}.

In looking back over the fantastic progress made in the last couple of years, and extrapolating into the future,
it seems probable that these, and other proposed projects, will have a fair chance to discover a WIMP 
signature within the present decade. 
In conjunction with indirect searches and accelerator production of new particles at the weak scale, 
they will allow to reveal the detailed properties of WIMPs, such as their mass, spin and couplings 
to ordinary matter, and shed light on their velocity and spatial distribution in our galactic halo.

\begin{acknowledgments}
The author wishes to thank the organizers for the invitation to a very stimulating 
and enjoyable symposium.

\end{acknowledgments}

\bigskip 

\end{document}